\documentclass[showpacs,aps,prl,twocolumn,superscriptaddress]{revtex4-1}

\pdfoutput=1

\usepackage{amsmath,amsfonts,amssymb,bm,graphicx,hyperref,color}

\begin{document}

\title{Comment on ``Is there a breakdown of the Stokes-Einstein relation in Kinetically Constrained Models at low temperature?'' by O. Blondel and C. Toninelli, arXiv:1307.1651 }



\maketitle

The answer to the question posed by Blondel and Toninelli in their title to Ref.\ \cite{Blondel2013} is clearly yes.  See Fig.\ 1.  The kinetically constrained model (KCM) considered in that figure is the East model \cite{Garrahan2010} coupled to a diffusing probe particle, a model first introduced in Ref.\ \cite{Jung2004}.   It is the model analyzed in Ref.\ \cite{Blondel2013}.  

The data presented in Fig.\ 1 extends that of Ref.\ \cite{Jung2004} by six orders of magnitude in relaxation time.  The range of conditions considered in Fig.\ 1 is the range of variation in $\tau$ that is accessible to reversible glass-forming melts.  For that range, the graphed results can be fit with a fractional Stokes-Einstein (SE) relationship, $D \propto \tau^{-\xi}$ with $\xi \approx 0.77$.  The value of the exponent $\xi$ is consistent with those used to fit experimental data \cite{Ediger}, and it is consistent with value proposed in Ref.\ \cite{Jung2004}.

The usual SE relationship is the mean-field result, $D \propto 1/\tau$.  The breakdown of this relationship, termed ``decoupling,'' is physically significant because it implies fluctuation dominance.  It is therefore important for the validity and applicability of the model that it exhibits decoupling consistent with experiment.

Blondel and Toninelli show that fractional SE behavior cannot be the result of the model in the limit of very low temperature \cite{Blondel2013}.  Specifically, they derive $q^2 \leqslant D\tau \leqslant 1/q^\alpha$, where $\alpha > 0$ and $q$ is the equilibrium concentration of excited sites, $q = (1 + e^{1/T})^{-1}$. This excludes fractional SE as $T \to 0$  because $\tau$ grows faster than any power of $1/q$ upon lowering temperature $T$. 
An acceptable asymptotic form is $D\tau \sim 1/q^\alpha$, for which Blondel and Toninelli suggest $\alpha = 2$.
This purported value of $\alpha$ is inconsistent with our numerical data.  The best fit, shown in the bottom panel of Fig.\ 1, gives instead $\alpha \approx 1.5$ or $1.6$, and a yet smaller exponent would be needed for a best fit of the lower temperature data.  

Whatever form is used to fit the data, the East model coupled to a diffusing probe exhibits significant decoupling, with violation of the SE relationship growing by orders of magnitude as temperature is lowered.

\vspace{2mm}

\noindent
YounJoon Jung$^{1}$, Soree Kim$^{1}$, Juan P. Garrahan$^{2}$, David Chandler$^{3}$ \\ 
{\small \em $^{1}$Department of Chemistry, Seoul National University, Seoul 151-747, Republic of Korea} \\
{\small \em $^{2}$School of Physics and Astronomy, University of Nottingham, Nottingham, NG7 2RD, UK} \\
{\small \em $^{3}$Department of Chemistry, University of California, Berkeley, California 94720, USA}

\bigskip

\bigskip

\begin{figure}[th!]
\includegraphics[width=.95\columnwidth]{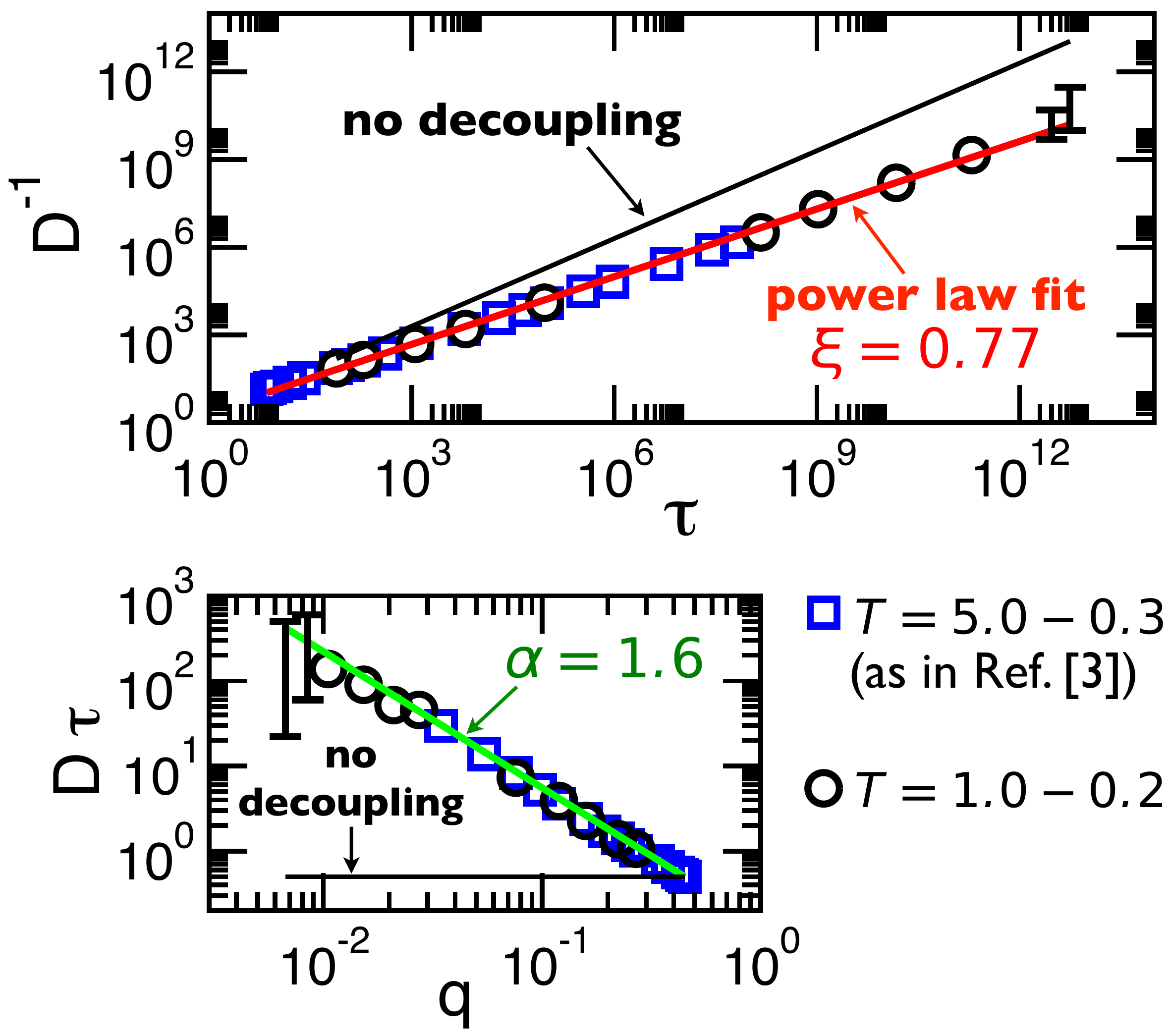}
\caption{Diffusing probe coupled to the East model, as in Ref.\ \cite{Jung2004}.  The top panel shows the inverse of the diffusion constant of the probe, $D^{-1}$, as a function of relaxation time $\tau$ (i.e., the average persistence time).  Blue squares are the original data \cite{Jung2004} (temperatures $5.0 \geqslant T\geqslant 0.3$).  Black circles are the new data ($1.0 \geqslant T \geqslant 0.2$).  The red line is a fit $D \sim \tau^{-\xi}$  over the whole range, and the black line the case of no SE breakdown.  Pronounced decoupling is obvious.  The bottom panel is a test of the $D \tau \sim q^{-\alpha}$ scaling proposed in \cite{Blondel2013}, with $q = (1 + e^{1/T})^{-1}$.  A best fit to the full range of data yields $\alpha \approx 1.6$ (rather than $2$, as suggested in Ref.\ \cite{Blondel2013}).  (For data collection, system sizes ranged from $L=10^{3}$ to $10^{6}$, and averages from $10^{4}$ to $10^{6}$ probe trajectories. Error estimates are generally smaller than symbols.  The exception is for $T=0.21$ and $0.2$.)}
\end{figure}

\end{document}